# Complex field measurement in a single pixel hybrid correlation holography

Ziyang Chen[1], Darshika Singh[2,1], Rakesh Kumar Singh[3,1,*], Jixiong Pu[1]

[1]*College of Information Science and Engineering, Fujian Provincial Key Laboratory of Light Propagation and Transformation, Huaqiao University, Xiamen, Fujian 361021, China*
[2]*Applied and Adaptive Optics Lab, Department of Physics, Indian Institute of Space Science and Technology (IIST), Valiamala, Trivandrum, India*
[3]*Department of Physics, Indian Institute of Technology (BHU), Varanasi, 221003, Uttar Pradesh, India*

*Corresponding author: krakeshsingh.phy@iitbhu.ac.in*

**We propose a new scheme for recovery of complex valued object in a single- pixel hybrid correlation holography. Idea is to generate an intensity correlation hologram by correlation of the intensity of a single pixel detector and digitally propagated two dimensional intensity patterns. The proposed scheme with theoretical basis is described to reconstruct the objects from a single pixel detector. An experimental arrangement is proposed and as a first step to realize the technique, we have simulated the experimental modelfor imaging of two different complex objects.**

Holography allows recording and reconstruction of the complex field. With optical recording and digital reconstruction in the holography, known as digital holography (DH), the technique has been attracting significant interests in quantitative phase measurement. Different DH schemes have been proposed to record and reconstruct the objects [1-5]. The phase imaging is essential to examine the complex valued object for various reasons. However, when the object is obscured by the scattering medium or illuminated by a random light, recovery of amplitude and phase information of the non-stochastic object is a challenging task. Such issues have been matter of investigations since long and several techniques have been proposed such as adaptive optics [6], phase conjugation [7, 8], correlation optics [9] etc.

Correlation techniques like ghost imaging and ghost diffraction have drawn significant interests [10-12]. This technique reconstructs the object from correlation of the intensity fluctuations from two channels, namely, the test and the reference. The test arm contains the object and equipped with a bucket detector that collects the light without any spatial resolution. The reference arm needs a space-resolving optical detection. The standard ghost imaging and diffraction provide image of the amplitude objects. A devised Young's interferometer has been utilized to measure the field correlation function to reconstruct the 1D complex valued object in a ghost diffraction scheme [13]. The diffraction pattern of a pure phase object with the incoherent light source has been discussed in Ref. [14].Ghost diffraction imaging of a pure phase objects with intensity correlation has also been described [15-19].Recently, significant interests have emerged on various computational correlation techniques such as computational ghost imaging [20, 21] and single-pixel imaging with random field illumination [22-24]. New methods to retrieve the complex information under single pixel detection scheme have also been proposed by structured mask illumination with phase shifting [25-27].

On the other hand, over the past few years correlation holography (CH) has emerged as a promising approach [28, 29].The CH reconstructs the object as a distribution of the two point correlation of the random field. Recently a new method, a combination of optical and computational channels, to reconstruct the object from a single pixel detector is proposed and referred as hybrid correlation holography (HCH) [30].The HCH provides reconstruction of amplitude and not adequate to recover the phase. In this letter, we propose a new approach to recover the complex valued object in a HCH framework.

Before going into theoretical basis of the proposed technique, let us compare recovery of the complex object from cross-covariance of the intensity and its implementation into a single pixel detection scheme. Consider a setup shown in Fig. 1(a) as in Ref. [31]. A monochromatic light coming out of the laser is collimated by a lens $L_1$. The collimated beam splits into two arms by a beam splitter (BS1) and propagates through two rotating ground glass (RGG). The transmitted light from the BS1 is focused by a microscope objective at an off-axis of the diffuser RGG1 and the random field is represented as $E_1(r) = T_1(r)e^{i\phi_1(r)}$. The off-axis position of the light at the RGG1 is controlled by the steering mirror M1. The field reflected by the BS1 illuminates the transparency T and travels through a RGG2. The complex amplitude of the random light immediately after the transparency is represented by $E_2(r) = T_2 e^{i\phi_2(r)}$. The coherent random fields from both paths interfere and recorded at the detector plane.

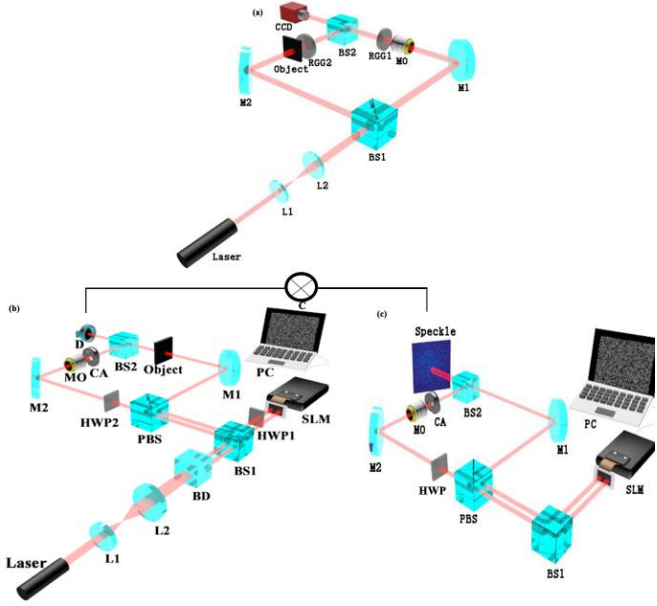

Fig. 1: (a) represents coherence waves interference setup with correlation of the intensity detected by the Charge Coupled Device (CCD); (b). Coherence waves interference in Hybrid correlation holography with single pixel detector setup: BS- beam splitter, SLM- spatial light modulator, c- correlator, D- single pixel detector (c) Architecture for digital propagation of the random fields without object and correlation with intensity of (b)

The instantaneous random intensity at the recording plane is represented as

$$I(u) = |E_1(u) + E_2(u)|^2 \qquad (1)$$
$$= \left| F\left\{T_1(r) e^{j\phi_1(r)}\right\} + F\left\{T_2(r) e^{j\phi_2(r)}\right\} \right|$$

where $F$ is the two dimensional Fourier transform (2D FT), $T_{i=1,2}$ represents the transmittance into path of random light and $\phi_i(r)$ is the random phase introduced by the rotating ground glass RGG. In the above expression, $u$ and $r$ represents two dimensional spatial coordinates at the detector and object planes respectively. For the Gaussian statistics, the cross-covariance of the intensity is expressed as [31-33]

$$\langle \Delta I(u) \Delta I(u + \Delta u) \rangle \propto |W_1(\Delta u) + W_2(\Delta u)|^2 \qquad (2)$$

Where the angular bracket <.> denotes the ensemble average and the random intensity variation from its mean intensity $\langle I(u) \rangle$ is $\Delta I(u) = I(u) - \langle I(u) \rangle$. Terms $W_1(\Delta u)$ and $W_2(\Delta u)$ represent the complex coherence of the random fields coming from RGG1 and RGG2 respectively. The complex coherence functions is represented as

$$W_i(\Delta u) = \langle E_i^*(u) E_i(u + \Delta u) \rangle \qquad (3)$$
$$= \int |T_i(r)|^2 \exp[-j2\pi r \cdot \Delta u] dr$$

Here making the use of the relation $\exp[j(\phi_i(r_1) - \phi_i(r_2))] = \delta(r_1 - r_2)$, Eq. (2) represents an interference of the coherence waves [32].

Let us turn attention to coherence waves interference in the HCH framework. There are significant differences between the proposed scheme and Fig. 1(a). In Fig. 1 (a), the cross-covariance is estimated for detected intensity by a two dimensional array of detectors, i.e. a charged coupled device (CCD). However, in the proposed new scheme we measure cross-covariance of the intensities coming from two channels, namely optical and digital which is represented as $\langle \Delta I_o(u) \Delta I_c(u + \Delta u) \rangle$, where $\Delta I_o(u)$ and $\Delta I_c(u)$ are intensity fluctuations in optical and digital channel respectively. In contrast to the CCD and RGG, proposed method uses a computer controlled spatial light modulator (SLM) to introduce random phases in the coherent beams and a single pixel detector. The proposed experimental arrangement is shown in Fig. 1 (b) and 1(c).Fig. 1(b) represents the optical implementation whereas Fig. 1(c) highlights digital propagation channel of the random intensity $I_c$. The detailed theoretical explanation, procedure and results are discussed as follows.

A monochromatic laser beam, after spatial filtering and collimated by lens combinations $L_1$ and $L_2$, enters into a beam displacer (BD) which generates the light into two parallel propagating orthogonal polarization components. These components propagate towards a SLM after passing through a beam splitter (BS1). The SLM is considered to modify only one of the orthogonal components of the light. Therefore, one of the orthogonal polarization components is flipped before and after reflection from the SLM by a half wave plate. The SLM introduces sequence of the random patterns into the spatially separated incoming beams which are further folded and directed by BS1 towards a Mach-Zhender (M-Zh) interferometer. Random fields entering into the M-Zh assembly is separated into the orthogonal components by a polarization beam splitter (PBS). A random pattern reflected from the PBS illuminates the object and propagates towards a single pixel detector after reflection from mirror M1 and passing through the BS2. On the other hand, a HWP2 is used to rotate the orthogonally polarized random field passing through the PBS in order to maintain the same polarization states in the interferometer. This random pattern is folded by mirror M2, focused and filtered by a microscope MO with pinhole CA. The transverse spatial separation of the random patterns at the SLM plane helps to introduce relative linear phase into the complex coherence function. The coherent random fields folded by mirror M1 and M2 passage through BS2 and interfere. Paths of the random fields in both the arms are matched in order to avoid any decorrelation due to difference in the longitudinal distances. The signal in the optical channel is proportional to the incident random intensity and represented as

$$I_o(u) = |E_o(u)|^2 = |E_o^1(u) + E_o^2(u)|^2 \qquad (4)$$

where $I_o(u)$ is non-averaged intensity corresponding to a random pattern over the SLM.

The intensity for the digital channel can be similarly expressed by using the Fourier propagation kernel. Propagation of the light in the digital channel is shown in Fig. 1(c). We have maintained a folded geometry of the beam in the digital channel in order to compare with its optical counterpart. The object transparency is removed during digital propagation of the random fields and estimation of the digital intensity.

For Gaussian random fields, the intensity correlation between two channels is expressed as

$$\langle I_c(u_1) I_o(u_2) \rangle = \langle E_c^*(u_1) E_c(u_1) \rangle \langle E_o^*(u_2) E_o(u_2) \rangle + \left| \langle E_c^*(u_1) E_o(u_2) \rangle \right|^2 \quad (5)$$

Considering statistical independence of the random fields introduced by the SLM, we express the second order correlation as

$$\langle E_c^*(u_1) E_o(u_2) \rangle = \langle E_c^{1*}(u_1) E_o^1(u_2) \rangle + \langle E_c^{2*}(u_1) E_o^2(u_2) \rangle \quad (6)$$

where $\langle E_c^i(u_1) E_o^i(u_2) \rangle = W_{co}^i(u_1, u_2)$ is complex coherence function and $i = 1, 2$. Term $i = 1$ and 2 represent the first and second arms of the M-Zh interferometer. The complex coherence function can be connected with the transparency as

$$W_{co}^i(u_1, u_2) = \left\langle \iint T_i(r_2) \exp\left[j(\phi(r_2) - \phi(r_1))\right] \times \exp\left[-j(u_2 \cdot r_2 - u_1 \cdot r_1)\right] dr_1 dr_2 \right\rangle \quad (7)$$

Eq. (7) represents the field correlation of a digitally evaluated field $E_c(u_1)$ with the field $E_o(u_2)$. Considering $u_2 = 0$ for single pixel detection and the incoherent source structure, Eq. (7) transforms to

$$W_{co}^i(u) = \int T_i(r) \exp(-ju.r) dr \quad (8)$$

Eq. (8) shows that the complex coherence is shaped by the Fourier transform of the transparency. Here term $T_i(r)$ can be real or complex and hence different from the van Cittert-Zernike condition discussed in Ref. [31] and Eq. (3). In order to measure the coherence in a single pixel scheme, we utilize Eq. (6) and write the cross-covariance as

$$\langle \Delta I_c(u) \Delta I_o(0) \rangle = |F\{T_1(r)\} + F\{T_2(r)\}|^2 \quad (9)$$

Eq. (9) is analogous to the holographic relation. To apply the off-axis holographic approach, we consider a reference coherence function $W_{co}^1(u) = F\{\delta(r - r_g)\}$ which provides a constant coherence with a linear phase. The complex coherence $W_{co}^2(u)$ dealing with object transparency can be recovered from other redundant factors by the Fourier analysis of Eq. (9).

Implementation of the proposed scheme is explained as follows. A SLM introduces independent random patterns at two spatially separated locations in the incoming beams as shown in Fig. 1b. The random phases are represented by $\phi_{nm}^i(M)$ on nm pixel, where M represents the number of random patterns. The random fields are independent and considered to follow $\langle e^{j\phi_{nm}^i(M)} \rangle = 0$ and $\langle e^{j[\phi_{nm}^i(M_2) - \phi_{jk}^i(M_1)]} \rangle = \delta_{jn} \delta_{km}$. A coherent field loaded with $e^{j\phi_{nm}^1}$ moves in a path without the object. Another coherent beam loaded with $e^{j\phi_{nm}^2}$ illuminates the object, and further propagates. The resultant random fields coming from both these arms are superimposed and intensity $I_o^n(0)$ is measured by a single-pixel detector. Here $n$ represents a random phase mask displayed on the SLM and ranges from 1 to M. The random fields displayed on the SLM are again digitally propagated using the Fourier transform kernels as in Fig. 1(c) and $I_c^n(u)$ is evaluated for the two dimensional coordinate $u$. The cross-covariance of the intensity from both channels is

$$\langle \Delta I_c(u) \Delta I_o(0) \rangle = \sum_{n=1}^{M} \Delta I_c^n(u) \Delta I_o^n(0) \quad (10)$$

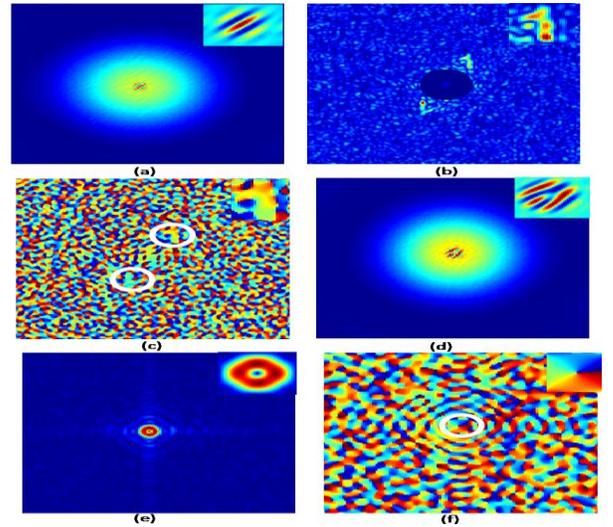

Fig.2: Recovery of the complex field in the single pixel HCH for two different transparencies. An off-axis hologram used as a transparency (a) cross-covariance of intensity (b) amplitude distribution of the coherence (c) phase distribution of the coherence; A spiral phase plate used as a transparency (e) cross-covariance of the intensity (f) reconstructed vortex amplitude (g) vortex phase. An interpolated portion of the results are also shown in the corner of each Figs for the demonstration purpose

Here M represents the number of random masks displayed over the SLM. In each realization, random phase mask of size 300X300 are introduced into the incident beam. In order to generate the desired reference coherence $W_{co}^1(u)$, we consider a point source placed at the off-axis of the MO. We use statistically independence of the random masks in the two paths, i.e. $\langle e^{j[\phi_{nm}^1(M) - \phi_{nm}^2(M)]} \rangle \approx 0$.

The results of the cross-covariance in the proposed hybrid correlation holography are shown in Fig. 2 for two different transparencies. Fig. 2(a) represents the cross covariance of the intensity corresponding to an off-axis hologram as a transparency. Presence of fringes in the cross-covariance confirms existence of interference as in Eq. (9). By Fourier analysis, the desired complex coherence is recovered and shown in Figs. 2(b) and 2(c) as amplitude and phase distributions. The strong central frequency content in Fig. 2(b), as in the case of the off-axis hologram, is suppressed to highlight the object. White circles in Fig. 2(c) highlight phase structure of the reconstructed object and its conjugate. For demonstration purpose, magnified and smoothened portion of the fringes is shown in the corner of each Figs, which is obtained by apply the 2D FT and interpolation of pixels in the signal domain.

In another case a spiral phase plate, which introduces a vortex phase structure, is used as a transparency. The light coming out of the spiral phase plate is represented as $r^m \exp(im\phi)$, where r and $\phi$ are transverse position and azimuthal coordinates respectively and m is topological charge of the vortex. For unit topological charge, the cross-covariance and reconstructed complex coherence are shown in Fig, 2(d)-2(f). Presence of a fork structure in the cross-covariance in the Fig, 2(d) confirms existence of a vortex structure in the object path. Using the Fourier analysis, the amplitude and phase structures of the vortex are retrieved as distribution of the coherence function. An accumulated phase variation around the heart of the vortex core, as represented by a white circle, confirms $2\pi$ phase and hence topological charge is unity. The reconstruction results shown in Fig.2 are obtained for M=100000 random samples. Reconstruction quality depends on the object aperture size and M values [30].

In order to examine effect of number of random masks M on the reconstruction quality, we have evaluated reconstruction efficiency and visibility for varying values of M and results are given in Table1. The visibility (V) is defined as the extent to which the reconstruction is distinguishable from the background noise. It is measured as the ratio of the average image intensity level in the region corresponding to the signal area to the average background intensity level. The reconstruction efficiency (R) is defined as the ratio of the measured power in the signal region of the image to the sum of this and the measured power in the background region [34].The reconstruction quality improves with an increase in the value of M. To perform quantitative evaluation of the reconstruction of the object with varying M values, we show reconstruction efficiency and visibility parameters in Table. 1.

Table -1. Visibility (V) and reconstruction efficiency (R)

| M=5000 | M=30000 | M=60000 | M=100000 |
|---|---|---|---|
| V=1.77 | V= 3.4 | V= 5.29 | V= 7.08 |
| R=0.639 | R=0.773 | R=0.841 | R=0.8763 |

In conclusion, a new technique to recover the complex field in a single pixel hybrid correlation holography is proposed. The basic principle of the technique is derived from making use of the interference pattern in the intensity correlation in a single pixel scheme. The experimental geometry and results of the simulated experimental models are presented for two different cases of the object. The proposed technique is find applications in the computational imaging, polarimetry and encryption.

**Funding:** Science Engineering Research Board (SERB) India Grant No EMR 2015/001613, National Natural Science Foundation of China (NSFC) under grant numbers 11674111, 61575070, 11750110426; Fujian Province Science Funds for Distinguished Young Scholar2018J06017; Natural Science Foundation of Fujian province under grant numbers 2017J01003.

**Acknowledgment**: RKS acknowledges seed grant from the IIT (BHU).